\documentclass[twocolumn,showpacs,prl]{revtex4}

\usepackage{graphicx}%
\usepackage{dcolumn}
\usepackage{amsmath}
\usepackage{latexsym}

\begin {document}

\title
{
A transition from river networks to scale-free networks
}
\author
{
A. K. Nandi and S. S. Manna
}
\affiliation
{
Satyendra Nath Bose National Centre for Basic Sciences
Block-JD, Sector-III, Salt Lake, Kolkata-700098, India
}
\begin{abstract}

      A spatial network is constructed on a two dimensional space where
   the nodes are geometrical points located at randomly distributed
   positions which are labeled sequentially in increasing order of one
   of their co-ordinates. Starting with $N$ such points the network is
   grown by including them one by one according to the serial number into 
   the growing network.
   The $t$-th point is attached to the $i$-th node of the network
   using the probability: $\pi_i(t) \sim k_i(t)\ell_{ti}^{\alpha}$
   where $k_i(t)$ is the degree of the $i$-th node and $\ell_{ti}$ is the
   Euclidean distance between the points $t$ and $i$.
   Here $\alpha$ is a continuously tunable parameter and while for $\alpha=0$ 
   one gets the simple Barab\'asi-Albert network, the case for 
   $\alpha \to -\infty$ corresponds to the spatially continuous version of the
   well known Scheidegger's river network problem. The modulating parameter
   $\alpha$ is tuned to study the transition between the two different
   critical behaviors at a specific value $\alpha_c$ which we numerically estimate to be -2.

\end{abstract}

\maketitle

       Scale-free networks (SFN) are highly inhomogeneous with a power law
    decay of their nodal degree distributions signifying the absence
    of a characteristic value for the nodal degrees \cite {barabasi}. Extensive research
    over last several years revealed that such networks indeed
    occur in different real-world systems like protein interaction
    networks in Biology, Internet and World-wide web (WWW) in electronic communication
    systems, airport networks in public transport systems etc.
    \cite {barabasireview, Dorogovtsev, Romuldo}.
    On the other hand, river networks are relatively simple spatial
    networks which were being studied over last several decades
    from the geological point of view. During the last decade or so
    physicists have also studied properties of river networks with
    many different simple model networks mainly from the interests
    generated about their fractal properties, a popular topic of
    critical phenomena \cite {Rinaldo}.

       In this paper we report our study of an weighted spatial network
    on the two-dimensional plane. The weight of a link is evidently the
    Euclidean length of the link. Tuning a parameter which modulates the
    strength of the contribution of the link weight in the attachment
    probability, we are able to obtain networks similar to the directed
    river network model in one limit. On the other limit of the parameter
    we obtain scale-free networks. The transition point for the crossover
    between the two types of behaviors is studied.

       The simplest river network model on a lattice is the Scheidegger's
    river network with a directional bias \cite {Scheidegger}. This is simply
    described on an oriented square lattice: Each lattice site is associated
    with an outgoing arrow representing the direction of the flow vector
    from that site. Only two possible choices for this arrow are possible:
    it may direct to the lower left lattice site or to the lower right.
    An independent and uncorrelated assignment of an arrow from each site
    results a `Directed Spanning Tree' (DST) network \cite {network,Rahul}. 
    Such networks are characterized
    mainly by the critical exponents associated with the distributions of the 
    river basin area as well as the
    length of the longest river at each site. The set of associated exponents
    constitute the universality class of the Scheidegger's river network
    which are different from the similar exponents of the isotropic
    spanning tree networks \cite {Manna-Dhar-Majum}.

      On the other hand while studying the scale-free properties of different
    real-world networks Barab\'asi and Albert (BA) argued that there is a `Rich get richer'
    mechanism in-built with the growth process of every SFN \cite {barabasi}. They proposed
    a model of generating scale-free networks where new nodes are introduced to the
    growing network at a rate of one per unit time step which are connected
    with the growing network with $m$ distinct links with a probability
    proportional to the individual nodal degree: $\pi_i(t) \propto k_i(t)$
    \cite {barabasi}. Also there are some other directed scale-free networks whose
    links are meaningful only when there is a connection from one end to
    the other but not along the reverse direction, e.g., the
    World-wide web \cite {web}, the phone-call network \cite {phone}
    and the citation network \cite {Redner} etc.

      Real-world networks whose nodes are geographically located in different
   positions on a two-dimensional Euclidean space are very important in their
   own right. For example the electrical networks in power transmissions,
   railway or postal networks in transport networks are few of the very well known
   spatial networks. Research over last few years have also revealed that
   two very important spatial networks like the Internet \cite {Faloutsos,Pastor,Yook} 
   and the Airport networks \cite {Guimera,Barrat} have scale-free structures. 

      Weighted networks are those whose links are
   associated with non-uniform weights $w_{ij}$. Therefore the spatial networks are by definition
   weighted networks whose link weights are the Euclidean lengths of the links.
   For a weighted network one can define the strength of a node $i$ as the
   total sum of the weights $s_i = \Sigma_jw_{ij}$. How the average nodal strength $\langle s(k) \rangle$
   varies with the degree $k$, i.e., $\langle s(k) \rangle \propto k^{\beta}$ is also
   a non-trivial question. Non-linear strength-degree relations have been observed for the
   Internet as well as the Airport networks. In this context a detailed knowledge of link length
   distribution is also important e.g., in the study of Internet's topological structure
   for designing efficient routing protocols and modeling Internet
   traffic. Early studies like the Waxman model describes the Internet with exponentially
   decaying link length distribution \cite {Waxman}.
   Yook et. al. observed that nodes of the router level network maps of North America
   are distributed on a fractal set and the link length distribution is inversely
   proportional to the link lengths \cite {Yook}. A number of model networks on the
   Euclidean space have been studied in different contexts
   \cite {Rozenfeld,Manna-Sen}.

%---------------------------------------------------------------------------
\begin{figure}[top]
\begin{center}
\includegraphics[width=7.0cm]{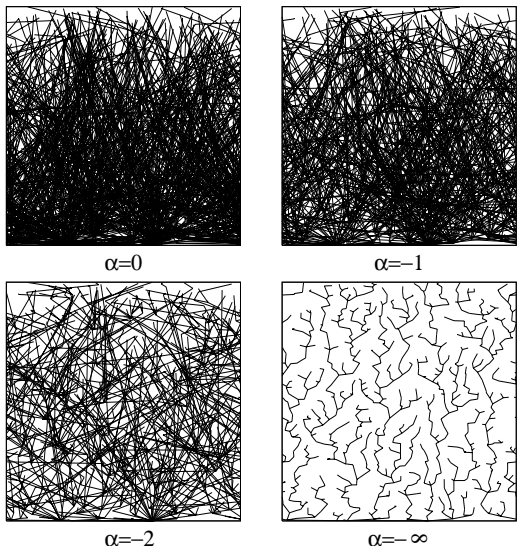}
\end{center}
\caption{
Pictures of the network generated for the four different values of the 
modulating parameter $\alpha$ using the same set of $N=1024$ randomly
distributed points within the unit square box. Link lengths are large
for $\alpha=0$ which gradually shrink in length as $\alpha$ decreases
and the network becomes the directed minimal spanning tree at $\alpha=-\infty$.
}
\end{figure}
%---------------------------------------------------------------------------

      We consider here a stack of nodes dropped one by one on a substrate with
   increasing vertical co-ordinates. Each node is connected randomly with a specific
   link length dependent probability of attachment to a node of the already grown
   stack.

      A network of $N$ nodes is grown within an unit square box on the two-dimensional
   $x-y$ plane. Nodes are represented by $N$ points selected at random positions
   ${(x_i,y_i)}, i=1,N$ by generating their values from an uniform probability distribution 
   $\{0,1\}$. The first point is placed by hand at the bottom of the box with $y_1=0$. All other 
   points are assigned serial numbers in increasing order of their $y$-coordinates: 
   $y_1 < y_2 < y_3 ... < y_N$. We use the geometry of a cylinder i.e., impose the open 
   boundary condition along the $y$-direction but the periodic boundary condition along 
   the $x$-direction. This implies that the space is continuous along the $x$-direction 
   and any node very close to the $x=1$ line may have a right neighbor inside the box 
   and very close to the line $x=0$ and vice-versa.

      To start with we assume that the first node at the bottom of the
   box has a `pseudo' degree $k_1=1$. Then the nodes from 2 to $N$ are
   connected to the network by one link each. The time $t$ measures
   the growth of the network by the number of nodes. The $t$-th node is then linked to
   the growing network with an attachment probability
   \begin {equation}
   \pi_i(t) \propto k_i(t) \ell_{ti}^{\alpha}
   \end {equation}
   where $\ell_{ti}$ denotes the Euclidean distance between the $t$-th and the $i$-th
   nodes maintaining the periodic boundary condition.
   This implies that the attachment probability has two competing factors. The
   linear dependence on the degree $k_i(t)$ enhances the probability of
   connection to a higher degree node where as the factor $\ell^{\alpha}$ reduces
   the probability of selection when $\alpha > 0$ and enhances when $\alpha < 0$.
   The special case of $\alpha=0$ the attachment probability in Eqn. (1) clearly
   corresponds to the Barab\'asi-Albert model. 
   We now discuss the properties of the network by continuously tuning the parameter
   $\alpha$ through its accessible range.

%---------------------------------------------------------------------------
\begin{figure}[top]
\begin{center}
\includegraphics[width=6.0cm]{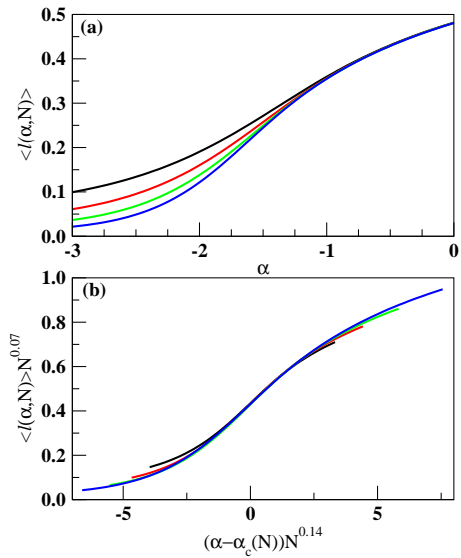}
\end{center}
\caption{(Color online)
The average length of a link $\langle \ell (\alpha,N) \rangle$ has been
plotted in (a) with the continuously tunable parameter $\alpha$ for different
values of the system sizes: $N=2^8, 2^{10}, 2^{12}$ and $2^{14}$. $N$ increases
from top to bottom.
In (b) a scaling is shown with the system size dependent critical value
of $\alpha_c$ which approaches -2 as $N \to \infty$.
}
\end{figure}
%---------------------------------------------------------------------------

   In the limit of $\alpha \to -\infty$ every node connects its nearest node $i$ in
   the downward direction corresponding to the smallest value of the link length $\ell$ 
   with probability one irrespective of its degree $k_i$ and therefore the probability
   of attachment to any other node is identically zero. This link may be directed 
   either to the left or to the right depending on the position of the nearest node (Fig. 1).

Let us first study the 
first neighbor distance distribution. Consider an arbitrary point P at an arbitrary 
position. The probability that its first neighbor is positioned on the semi-annular
ring within $r$ and $r+dr$ in the downward direction (which can be done in $N-1$ 
different ways) and all other $N-2$ points are at distances larger than $r+dr$ is:
\begin {equation}
{\rm Prob}(r,N)dr = (N-1) \hspace*{0.1 cm} \pi r dr \hspace*{0.1 cm} (1-\pi r^2/2)^{N-2}.
\end {equation}
In the limit of $N \to \infty$ it can be approximated that $N-1 \approx N-2 \approx N$ 
and since the average area per point decreases as $1/N$, $\pi r^2/2$ is very 
small compared to 1. Therefore $(1-\pi r^2/2)^{N-2}$ is approximated as 
$\exp (-\pi N r^2/2)$. In the limit of $N \to \infty$ the probability 
density distribution is therefore:
\begin {equation}
{\rm Prob}(r,N) = \pi N r \hspace*{0.1 cm} \exp (-\pi N r^2/2)
\end {equation}
or in the scaling form:
\begin {eqnarray}
\frac {1}{\sqrt{N}}{\rm Prob}(r) & = & [\pi r\sqrt {N}] \hspace*{0.1 cm} \exp (-\pi [r\sqrt {N}]^2/2) = {\cal G}(x)
\end {eqnarray}
where the scaling function ${\cal G}(x) = \pi x \exp(-\pi x^2/2)$ and $x=r\sqrt {N}$.
Numerical results for the link length distribution of different system sizes verifies this
distribution very accurately.

%---------------------------------------------------------------------------
\begin{figure}[top]
\begin{center}
\includegraphics[width=7.0cm]{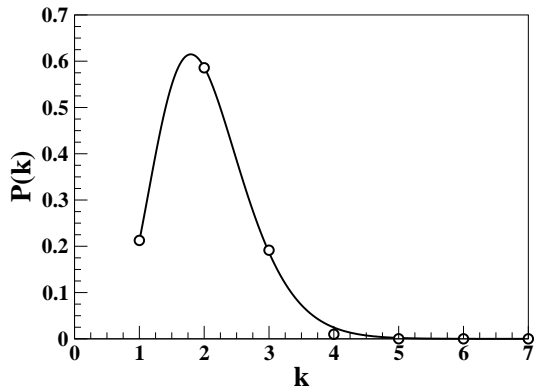}
\end{center}
\caption{
The degree probability distribution $P(k)$ vs. $k$ of the network in the
limit of $\alpha \to -\infty$ and for $N=2^{18}$. The solid line is a fit
and its functional form is given in Eqn. (7). 
}
\end{figure}
%---------------------------------------------------------------------------

   The typical length $\langle \ell (\alpha,N) \rangle$ of a link for a network of size $N$
   and generated with a specific value of the parameter $\alpha$ is estimated by averaging
   the link length over all $N-1$ links of a network as well as over many independent configurations.
   For this purpose one can define a total cost function ${\cal C}(\alpha,N)$ of the network 
   which is the total length of all the links:
\begin {eqnarray}
{\cal C}(\alpha,N) = \Sigma_{j=1}^{N-1}\ell_j \hspace*{0.1 cm} {\rm and} \hspace*{0.1 cm} 
\langle \ell (\alpha,N) \rangle = {\cal C}(\alpha,N)/(N-1).
\end {eqnarray}
   In Fig. 2(a) we show the variation of the average link length with $\alpha$. 
   The $\langle \ell (\alpha,N) \rangle \to 0$ as $\alpha \to -\infty$ and gradually 
   increases with increasing $\alpha$. Around $\alpha_c = -2$ the $\langle \ell (\alpha,N) \rangle$
   increases very fast and finally approaches unity as $\alpha \to +\infty$. The steep growth
   around $\alpha_c$ becomes increasingly sharper with increasing system size. 
   For very large networks $N \to \infty$ it appears that for $\alpha < \alpha_c$
   $\langle \ell (\alpha,N) \rangle \to 0$ and for $\alpha > \alpha_c$ it approaches a finite value. 
   Such a system size dependence is quantified by a finite-size scaling of this plot 
   as shown in Fig. 2(b). The data collapse shows that 
   \begin {equation}
   \langle \ell (\alpha,N) \rangle \propto N^{-0.07} {\cal G}[(\alpha-\alpha_c(N))N^{0.14}].
   \end {equation}
   The critical values of $\alpha_c(N)$ for a system of size $N$ is located at the value of $\alpha$
   where $\langle \ell (\alpha,N) \rangle$ increased most rapidly with $\alpha$. The values of $\alpha_c(N)$ 
   so obtained are $-1.37$, -1.46. -1.54 and -1.60 for $N = 2^8, 2^{10}, 2^{12}$ and
   $2^{14}$ and are extrapolated with $N^{-0.1}$ to get $\alpha_c = \alpha_c(\infty) =
   -2.0 \pm 0.10$. Similar results of $\alpha_c=-2$ have been obtained in \cite {Kleinberg,Roberson}. 

%---------------------------------------------------------------------------
\begin{figure}[top]
\begin{center}
\includegraphics[width=6.0cm]{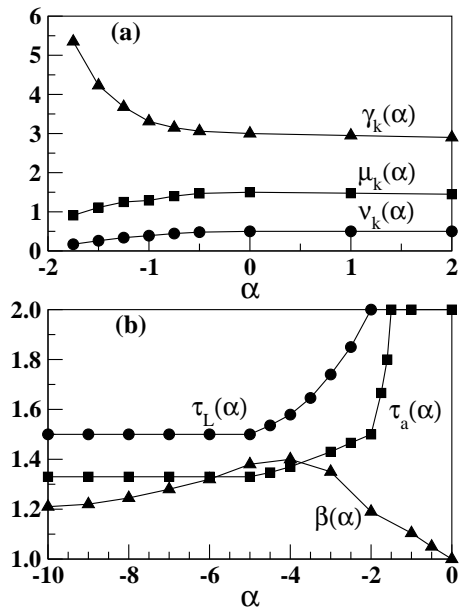}
\end{center}
\caption{
(a) Variation of the scaling exponents $\nu_k(\alpha)$, $\mu_k(\alpha)$ and $\gamma_k(\alpha)$
characterizing the degree distribution in Eqn. (8) with the modulating parameter $\alpha$.
(b) Variation of the exponents $\tau_a(\alpha)$, $\tau_L(\alpha)$
and $\beta(\alpha)$ with $\alpha$.
}
\end{figure}
%---------------------------------------------------------------------------

      In the limit of $\alpha \to -\infty$ and $N \to \infty$, the fractions of nodes with
   degrees 1, 2 and 3 are found to be 0.213, 0.586 and 0.192 respectively and decreases very fast for
   higher degree values. The whole distribution fits nicely to a sharp Gamma distribution as (Fig. 3):
   \begin {equation}
   P(k) \propto k^{7.5}\exp(-4.2k).
   \end {equation}
   In the range $\alpha > \alpha_c$ we observed that the network has a scale-free structure.
   For large system sizes the degree distribution follows a power law like $P(k) \propto k^{-\gamma}$
   but for finite systems a finite-size scaling seems to work well:
   \begin {equation}
   {\rm Prob}(k,N) \propto N^{-\mu_k}{\cal F}_k(k/N^{\nu_k}).
   \end {equation}
   The scaling function ${\cal F}_k(x) \sim x^{-\gamma_k}$ for $x \to 0$ and decreases faster
   than a power law for $x >> 1$ so that, $\gamma_k = \mu_k/\nu_k$. For a range of
   $\alpha$ values the scaling exponents are measured and it is observed that all three 
   exponents $\gamma_k, \mu_k$ and $\nu_k$ are dependent on the value of $\alpha$ (Fig. 4(a)).

      For the river network problems the size of the drainage area is a popular quantity 
   to measure. The amount of water that flows out of a node of the river network
   is proportional to the area whose water is drained out through this node. 
   On a tree network the drainage area $a_i$ is defined at every node $i$ and is measured 
   by the number of nodes supported by $i$ on the tree network. A well known
   recursion relation for $a_i$ is:
   $a_i = \Sigma_j a_j\delta_{ij} + 1$
   where the dummy index $j$ runs over the neighboring nodes of $i$ and $\delta_{ij}=1$
   if the flow direction is from $j$ to $i$, otherwise it is zero. The probability 
   distribution ${\rm Prob}(a)$ of the drainage areas
   is the probability that a randomly selected node has the area
   value $a$. It is known that for river networks this distribution
   has a power law variation: ${\rm Prob}(a) \sim a^{-\tau_a}$ \cite {Rinaldo}.

%---------------------------------------------------------------------------
\begin{figure}[top]
\begin{center}
\includegraphics[width=6.0cm]{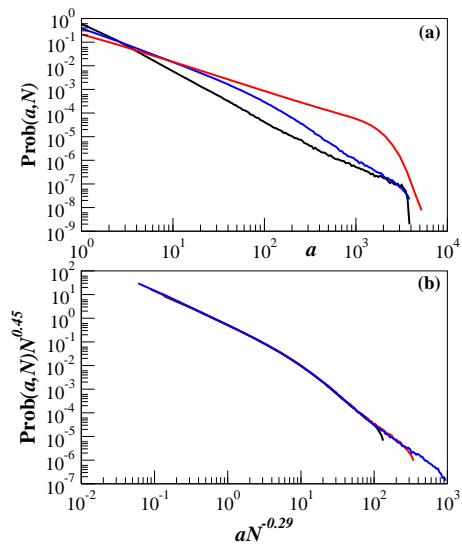}
\end{center}
\caption{(Color online)
Plot of the river basin area $a$ distribution in our model for different
network sizes $N$. In (a) ${\rm Prob}(a,N)$ has been shown for three different
values of $\alpha$: $\alpha=-\infty$ at the top, $\alpha=\alpha_c$ at the
middle and $\alpha=0$ at the bottom. In (b) a finite size scaling of
${\rm Prob}(a,N)$ with $N$ has been shown for $\alpha=\alpha_c$ for
$N=2^8, 2^9$ and $2^{10}$.
}
\end{figure}
%---------------------------------------------------------------------------

   The drainage area distribution is measured first in the limit of $\alpha \to -\infty$
   for our networks of different sizes $N = 2^{12}, 2^{14}$ and $2^{16}$. Direct measurement 
   of the slopes of double logarithmic plots of ${\rm Prob}(a,N)$ vs. $a$ gives values of the 
   exponent $\tau \approx 1.33$ which varied little with the system size (Fig. 5(a)). 
   This estimate is consistent with that obtained from the finite size scaling analysis.
   An excellent scaling of the data over different system sizes is obtained as:
   \begin {equation}
   {\rm Prob}(a,N) \propto N^{-\mu_a}{\cal F}(a/N^{\nu_a})
   \end {equation}
   and the exponent $\tau_a = \mu_a/\nu_a$.
   In the limit of $\alpha \to -\infty$ we estimated $\mu_a(-\infty) \approx 1.00$ and 
   $\nu_a(-\infty) \approx 0.75$ giving a value for the exponent $\tau_a(-\infty) \approx 1.33$.
   These values are very much consistent with the same exponents of Scheidegger's river network model
   where $\tau_a=4/3$ is known exactly \cite {Scheidegger}.
   Similarly for $\alpha=0$ we could reproduce the known values of $\mu_a(0) \approx 2.00$ and
   $\nu_a(0) \approx 1.00$ with $\tau_a(0) \approx 2.00$ as obtained in \cite {Redner}. Finally 
   we measured the same distribution 
   at the transition point $\alpha_c=-2$ and obtained $\mu_a(-2) \approx 0.45$ and
   $\nu_a(2) \approx 0.29$ with $\tau_a(-2) \approx 1.50$ (Fig. 5(b)).

   Another quantity of interest is the length $L_i$ of the longest up-stream meeting
   at the node $i$. It's magnitude is the number of links on the longest path terminating
   at $i$. ${\rm Prob}(L)$ therefore denotes the probability that an arbitrarily selected node $i$ has
   $L_i=L$. Given a network of size $N$, $L$ values
   are measured at every node and then the data is sampled over many uncorrelated
   network configurations and for different network sizes. 
   A similar finite size scaling form like
   ${\rm Prob}(L,N) \propto N^{-\mu_L}{\cal G}(L/N^{\nu_L})$
   works here as well. We obtain 
   $\mu_L(-\infty) \approx 0.75 $, $\nu_L(-\infty) \approx 0.50$, $\tau_L(-\infty) = 1.5$ and
   $\mu_L(2) \approx 0.2$, $\nu_L(2) \approx 0.1$, $\tau_L(2) \approx 2.0$.

   Finally, we studied the strength-degree relation in our model. Here the weight 
   of a link is the length of the link $\ell_{ij}$ and therefore
   the strength of the node $i$ is $s_i = \Sigma_j\ell_{ij}$. 
   The strength $\langle s(k) \rangle$ per node averaged over all nodes of degree $k$ 
   of the network as well as over many independent realizations varies with the degree $k$ as:
   $\langle s(k,\alpha) \rangle \propto k^{\beta(\alpha)}$. Numerically we observe that 
   the exponent $\beta(\alpha)$ varies with the tuning parameter $\alpha$. 
   Fig. 4(b) summarizes the variation of the three exponents $\tau_a(\alpha)$, $\tau_L(\alpha)$
   and $\beta(\alpha)$ within the range of $\alpha$ varying between -10 and 0. For
   $\alpha \le -5$, $\tau_a(\alpha)$ and $\tau_L(\alpha)$ values coincide with their values
   at $\alpha=-\infty$. Between $-5 < \alpha \le -2$, $\tau_L(\alpha)$ slowly increases to 2
   and $\tau_a(\alpha)$ increases to 1.5. 

   To conclude, we have defined and studied a network embedded in the Euclidean space.
   Random distribution of nodes are sequentially numbered in increasing heights and
   the degree dependent attachment probability is modulated by the $\alpha$-th power
   of the link length. This continuously tunable parameter $\alpha$ interpolates between 
   the Scheidegger's river network and the Barabasi-Albert Scale-free network. 
   It appears that there exists a critical value $\alpha_c$ such that 
   for $\alpha < \alpha_c$ the critical behavior of the network is like the Scheidegger's
   river network, whereas for $\alpha > \alpha_c$ critical exponents are indistinguishable
   from those of ordinary BA network. Our numerical study indicates $\alpha_c$ is likely
   to be -2.

   Discussion with G. Mukherjee and K. Bhattacharya are thankfully acknowledged.

\leftline {Electronic Address: manna@bose.res.in}


\begin{thebibliography}{90}

\bibitem {barabasi} A.-L. Barab\'asi and R. Albert, Science, {\bf 286}, 509 (1999).

\bibitem {barabasireview} R. Albert and A.-L. Barab\'asi, Rev. Mod. Phys. {\bf 74}, 47 (2002).

\bibitem {Dorogovtsev} S. N. Dorogovtsev and J. F. F. Mendes, {\it Evolution of Networks},
Oxford University Press, 2003; M. E. J. Newman, SIAM Review 45, 167 (2003).

\bibitem {Romuldo} R. Pastor-Satorras and A. Vespignani, 
{\it Evolution and Structure of the Internet, A Statistical Physics Approach},
Cambridge University Press, 2004.

\bibitem {Rinaldo} I. Rodríguez-Iturbe, A. Rinaldo, 
{\it Fractal River Basins: Chance and Self-Organization},
Cambridge University Press, 2001.

\bibitem {Scheidegger} A.E. Scheidegger, Geol. Soc. Am. Bull. {\bf 72}, 37 (1961); 
Water Resour. Res. {\bf 4}, 167 (1968).

\bibitem {network} F. Harary, {\it Graph Theory}, Addison-Wesley Publishing 
                Company, Inc., Reading, Mass., 1969.

\bibitem {Rahul} A. G. Bhatt and R. Roy, Appl. Probab. {\bf 36}, 19 (2004).

\bibitem {Manna-Dhar-Majum} S. S. Manna, D. Dhar and S. N. Majumdar, Phys. Rev. A. {\bf 46} R4471 (1992).

\bibitem {web} S. Lawrence and C. L. Giles, Science, {\bf 280}, 98 (1998);
               Nature, {\bf 400}, 107 (1999), R. Albert, H. Jeong and A.-L. Barab\'asi,
               Nature, {\bf 401}, 130 (1999).

\bibitem {phone} W. Aiello, F. Chung and L. Lu in Proc. 32-nd ACM Symp. Theor. Comp. (2000).

\bibitem{Redner} P. L. Krapivsky and S. Redner, Phys. Rev. E. {\bf 63}, 066123 (2001).

\bibitem {Faloutsos} M. Faloutsos, P. Faloutsos and C. Faloutsos, Proc.
               ACM SIGCOMM, Comput. Commun. Rev., {\bf 29}, 251 (1999).

\bibitem {Pastor} R. Pastor-Satorras, A. Vazquez and A. Vespignani, Phys. Rev. Lett. {\bf 87}, 258701 (2001).

\bibitem {Yook} S. H. Yook, H. Jeong and A.-L. Barab\'asi, Proc. Natl. Acad. Sci. (USA) {\bf 99}, 13382 (2002).

\bibitem {Guimera} R. Guimera and L. A. N. Amaral, Eur. Phys. Jour. B, {\bf 38}, 381 (2004).

\bibitem {Barrat} A. Barrat, M. Barth\'el\'emy, R. Pastor-Satorras, A. Vespignani, Proc. Natl. Acad. Sci. (USA),
{\bf 101}, 3747 (2004). 

\bibitem {Waxman} B. Waxman, IEEE J. Selec. Areas Commun., SAC, {\bf 6}, 1617 (1988).

\bibitem{Rozenfeld} A. F. Rozenfeld, R. Cohen, D. b-Avraham and S. Havlin, Phys. Rev. Lett.
{\bf 89}, 218701 (2002).

\bibitem {Manna-Sen} S. S. Manna and P. Sen, Phys. Rev. E {\bf 66}, 066114 (2002).

\bibitem {Kleinberg} J. M. Kleinberg, Nature, {\bf 406}, 845 (2000).

\bibitem {Roberson} M. R. Roberson and D. ben-Avraham, Phys. Rev. E {\bf 74}, 017101 (2006).
\end{thebibliography}
\end {document}